\documentclass[aps,prb,superscriptaddress,amsfonts,amsmath,amssymb,showpacs,floatfix,reprint,longbibliography,nofootinbib]{revtex4-2}

\usepackage{natbib}
\usepackage[english]{babel}
\usepackage{letltxmacro}
\usepackage{latexsym}
\LetLtxMacro{\ORIGselectlanguage}{\selectlanguage}
\makeatletter
\DeclareRobustCommand{\selectlanguage}[1]{%
 \@ifundefined{alias@\string#1}
 {\ORIGselectlanguage{#1}}
 {\begingroup\edef\x{\endgroup
 \noexpand\ORIGselectlanguage{\@nameuse{alias@#1}}}\x}%
}
\newcommand{\definelanguagealias}[2]{%
 \@namedef{alias@#1}{#2}%
}
\makeatother
\definelanguagealias{en}{english}
\definelanguagealias{English}{english}
\usepackage{graphicx}
\usepackage{amsmath}
\usepackage{amsfonts}
\usepackage{amssymb}
\usepackage{bm}
\usepackage{color}
\usepackage[percent]{overpic}
\usepackage{soul} 
\usepackage{amssymb}
\usepackage{wasysym}
\usepackage{dsfont}
\usepackage{float}
\usepackage[mathscr]{euscript}
\usepackage{lipsum}
\usepackage{hyperref}
\hypersetup{
 pdfstartview={FitH},
 colorlinks=true,
 linkcolor=blue,
 citecolor=blue,
 filecolor=magenta,
 urlcolor=blue
}

\newcommand{\pdagger}{\phantom{\dagger}}

\usepackage{mathrsfs}
\graphicspath{{./figures/}}

\newcommand{\ipr}[0]{\mathrm{IPR}}

\makeatletter
\renewcommand{\p@subsection}{}
\renewcommand{\p@subsubsection}{}
\makeatother

\usepackage{xcolor}
\usepackage{mathtools}
\usepackage{tikz}

\usetikzlibrary{automata,positioning,backgrounds}
\usetikzlibrary{decorations.markings}
\usepackage{dsfont}

\begin{document}
\title{Localization in one-dimensional magnetic quasicrystals}
\author{Andrew Osborne}
\email{andrew.osborne@princeton.edu}
\affiliation{Department of Electrical and Computer Engineering, Princeton University}
\author{Rhine Samajdar}
\affiliation{Department of Electrical and Computer Engineering, Princeton University}
\email{rhine_samajdar@princeton.edu}
\author{Andrew A. Houck }
\affiliation{Department of Electrical and Computer Engineering, Princeton University}
\email{aahouck@princeton.edu}
\begin{abstract}
The celebrated Aubry-Andr\'e-Harper model undergoes a metal-insulator transition akin to Anderson localization due to an energetic competition between nearest-neighbor hopping and a quasiperiodic on-site potential.
In this work, we consider the problem of a one-dimensional lattice subject to a quasiperiodic external field, where the topology of the underlying graph mediates the competition between insulating and conducting phases. 
In particular, we present examples for which all dispersive eigenstates are critically localized, as well as models in which they are exponentially localized. 
We also demonstrate the coexistence of these phenomena with an extensive set of degenerate finitely supported eigenstates forming a flat band. 
Finally, we construct models with exactly solvable mobility edges arising only from graph structure. Going beyond these single-particle results, we show---using exact-diagonalization calculations of the bosonic dynamics---that, as in the conventional Aubry-Andr\'e-Harper case, on-site interactions drive localized eigenstates toward extendedness. Our results provide a framework for exploring how magnetic interference and interactions jointly
constrain transport in quasiperiodic lattices.
\end{abstract}

\maketitle

\section{Introduction}
In one-dimensional (1D) systems, arbitrarily weak disorder leads to exponential localization of single-particle eigenstates \cite{anderson,diffusion}. The question of whether that localization persists in the face of interactions is the subject of the many-body localization (MBL) literature \cite{Abanin_2019,D_Alessio_2016,Nandkishore_2015}, which is active and sprawling. On the other hand, periodic systems give rise to extended eigenstates characterized by smooth dispersion and extensive support. 

Quasicrystals are a natural intermediary between genuinely disordered systems and periodic ones \cite{quasicrystal}.
Instead of relying upon genuine randomness, 1D quasicrystals use ergodic maps to fill the role of disorder without relinquishing deterministic model structure. 
A 1D quasicrystal is a hopping model for which the underlying graph and the modulation of the coefficients thereupon are both symmetric under some translation, but in such a way that the periods of the lattice and the modulation never align, yielding a model without translation symmetry at all.
The most well-studied quasiperiodic system is 
the Aubry-Andr\'e-Harper (AAH) model \cite{PGHarper_1955,aubry1980analyticity}
\begin{equation}\label{eqn:aah}
    E \psi_n = t \psi_{n-1} + t \psi_{n+1} +\lambda \cos(\Phi_n) \psi_n
\end{equation}
with 
\begin{equation}
    \Phi_n = 2 \pi \beta n + \theta
\end{equation}
and irrational $\beta$.
The AAH model arises naturally from the Hofstadter problem on a square lattice in two dimensions (2D) after a partial Fourier transform \cite{oghof}, and it is also the subject of the Ten Martini problem~\cite{Avila}. 
To mathematicians and physicists alike, the AAH model has been a paradigmatic testbed for the study of localization and has inspired a variety of generalizations \cite{Dom_nguez_Castro_2019,Xu_2021,Gopalakrishnan_2017,Liu_2020,liu_phase_2018,Xiaoshankar}.

When $\beta$ is rational, all of the eigenstates of Eq.~\eqref{eqn:aah} are extended and behave as ordinary Bloch waves, albeit in perhaps a great number of bands. 
However, when $\beta$ is irrational,  the qualitative behavior of single-particle eigenstates depends on the relative strength of $\lambda $ and $t$. 
For $|\lambda/t|<2$, eigenstates are qualitatively similar to plane waves in their extensive support and continuous energy spectrum \cite{avila_ac_2008}, as is clear to see from the $\lambda/t \rightarrow 0$ limit, which recovers translation invariance. 
On the other hand, for $|\lambda/t|>2$, eigenstates all Anderson localize,\footnote{Rigorously, this holds for Diophantine $\beta$ (such as the inverse golden ratio used in our numerics) and almost every phase $\theta$.} in a manner exactly analogous to the behavior of a disordered 1D chain \cite{jitomirskaya_metal_1999}. 
These two behaviors meet at $|\lambda/t|=2$, where all eigenstates on the entire spectrum are critical \cite{Avila_2017}, decaying in space as a long-range power law rather than an exponential. 
Remarkably, the localization properties of every eigenstate of the AAH model are controlled only by $|\lambda/t|$ and are independent of energy.
Generalizations of the AAH model \cite{ganeshan_nearest_2015,Jiang_2025,wang_duality_2021,Xiaoshankar,PhysRevLett.131.176401,Wang_2023} have, on the other hand, been shown to have localization properties dependent upon energy. 
Although energy-dependent localization behavior should be regarded as typical, it is still nontrivial to construct systems with exact mobility edges, but the literature nonetheless provides a healthy set of examples. 

In the case of the 2D Hofstadter problem, a constant magnetic field applied to a 2D lattice (with irrational flux per plaquette in units of the flux quantum) fails to admit a periodic magnetic unit cell, and hence the Fourier transform along a direction in which the chosen gauge is translation invariant gives rise to an effective 1D model with a quasiperiodic on-site potential. 
In contrast, we begin with a translation-invariant 1D model and apply a linearly increasing perpendicular magnetic field. 
For some values of the magnetic field strength, the reduced flux on unit cells becomes incommensurately periodic with the lattice spacing: it is precisely this setup that we call a magnetic quasicrystal. 
Unlike in the case of 1D quasicrystals induced by the Hofstadter problem, magnetic quasicrystals have an energetic competition mediated not by on-site potentials but instead by Aharonov-Bohm interference between paths through space. 

As in disordered 1D models, disordered magnetic fields generally induce Anderson localization in 1D, apart from possible modes at zero energy that are protected by a chiral symmetry \cite{PhysRevB.59.13221}. 
In distinction to the case of genuinely periodic 1D chains, however, periodic (even constant) magnetic fields are capable of flattening all bands in a system at once \cite{Vidal_1998,osborne2025gluedtreelatticescompact}. 
In this work, we explore the interplay of lattice topology with quasicrystal physics mediated by incommensurately periodic magnetic fields.

In almost all of what follows, we will insist that quasi-1D models have uniform hopping weight, in order to isolate the role of lattice properties. 
Because there is generally only a single scale in the problem, the interplay of the topological properties and microscopic details of the underlying lattices fills the role of the ratio $\lambda/t$ from the AAH model.
Unsurprisingly, the space of conceivable quasi-1D lattices is very complicated; it is beyond the scope of this paper to classify all lattices according to their localization phenomenology under quasiperiodic fields. 
Instead, we present a suite of examples demonstrating a variety of phenomena.

First, we create an infinite family of models spanning from full-spectrum criticality to full-spectrum exponential localization away from zero energy.
These models are quasiperiodic extensions of glued tree lattices known to disconnect entirely under constant fields \cite{7nzr-14dw,osborne2025gluedtreelatticescompact}, and we thus call them glued tree quasicrystals. 
Glued tree quasicrystals give rise to singular Jacobi transfer matrices that escape the theorems of Refs.~\cite{avila_global_2009,jitomirskaya_one-dimensional_2023,jitomirskaya_arithmetic_2017,han_non-perturbative_2023,klein_anderson_2017}, so we rely upon numerics to study all but the simplest model in this family.
Such glued tree quasicrystals also generally have an extensively degenerate flat band that coexists with the standard quasicrystal trichotomy of localized, critical, and extended states.
Thereafter, we introduce another family of models that yield exact mobility edges between extended and localized states by using standard Schr\"odinger operator transfer matrix complexification techniques after the decimation of some sublattices.
Finally, we argue that on-site interactions drive eigenstates toward delocalization, as in the ordinary AAH case \cite{Mastropietro_2015,Doggen_2019}, investigating the dynamics of interacting bosons on the glued tree chains in the few-body regime.

\section{Quasicrystals}

We say that a lattice is (quasi) 1D if there exists an integer $D$ such that $\langle n |H|m\rangle = 0$ whenever $|n-m| > D$. 
This is a milder imposition than the requirement that a lattice be genuinely 1D, but this liberty is important on the grounds that a magnetic field applied to a lattice without cycles can be simply gauged away.\footnote{Hence, it immediately follows that an incommensurately periodic external field cannot localize a proper 1D chain. This is our first clue that lattice geometries are very important to the problem of localizing under incommensurate magnetic fields, although this example is admittedly trivial.}  
Provided that a 1D graph has a translation symmetry, it is always possible to find matrices $V$ and $T$ such that the tight-binding single-particle Hamiltonian is block tridiagonal with diagonal blocks $V$ and off-diagonal blocks $T$ (after enlarging the unit cell if necessary). Explicitly, if a lattice has $q > 0 $ sites per unit cell, we may write 
\begin{equation}
    H = \begin{pmatrix}
\ddots & \ddots &0 & & \\
\ddots & V & T &0 & \\
 & T^\dagger & V & T & \\
 & 0 & T^\dagger & V & \ddots \\
 & &0 & \ddots & \ddots
\end{pmatrix},
\end{equation}
where $T$ and $V$ are each $q \times q$ matrices.

Let us choose some 1D graph drawn on a strip and subject the strip to the perpendicular magnetic field $\mathbf{B}=\mathcal{B}x\hat{\mathbf{z}}$, where $\mathcal{B}$ is a constant field gradient.\footnote{Here, $\beta$ denotes the increase in flux through a reference plaquette under one cell translation, measured in flux quanta. We consider embeddings with commensurate plaquette areas, for which the one-frequency form below applies.}  
This field is linearly increasing in the direction of the strip's translation symmetry and constant in the other. 
We will refer to these models as 1D magnetic quasicrystals, which are still block tridiagonal without loss of generality, but the blocks generally have some dependence upon space:
\begin{equation}\label{eqn:blocktr}
    H = \begin{pmatrix}
\ddots & \ddots &0 & & \\
\ddots & V^{\pdagger}_{n-1}(\theta) & T^{\pdagger}_{n-1}(\theta) &0 & \\
 & T^\dagger_{n-1}(\theta) & V^{\pdagger}_n(\theta) & T^{\pdagger}_n(\theta) & \\
 & 0 & T^\dagger_n(\theta) & V^{\pdagger}_{n+1}(\theta) & \ddots \\
 & &0 & \ddots & \ddots
\end{pmatrix},
\end{equation}
where 
\begin{equation}
    \begin{aligned}
        T_n(\theta)&= T_n(\theta + P ), \\
        V_n(\theta)&= V_n(\theta + P ), \\
        T_{n+1}(\theta) &= T_n(\theta + 2 \pi \beta), \\ 
        V_{n+1}(\theta) &= V_n(\theta + 2 \pi \beta),
    \end{aligned}
\end{equation}
with $\beta$ irrational and $P=2\pi$ in a periodic gauge (note that $P=4\pi$ for the half-phase gauges displayed below).
We will refer to $\theta$ as a ``phase'' and $\beta$ as a ``frequency.'' Except where explicitly mentioned, we will require all frequencies to be irrational. 
Finally, we will also use the notation $\Phi_n \equiv 2 \pi \beta n + \theta$ throughout the paper. 
In order to exclude the already very broad set of 1D quasicrystals, we will focus on effects that arise because of the quasiperiodic Peierls phase and thus generally suppose that the entries of $V_n(\theta)$ and $T_n(\theta)$ are $0$ or $1$ in magnitude.

If $T_n(\theta)$ is constant with respect to $\theta$ and invertible, and 
\begin{equation}
    V_n(\theta) = 
    \lambda
    \begin{pmatrix}
       \cos( 2 \pi \beta n  + \theta  ) & 0 & \dots & \\ 
       0 & \cos( 2 \pi \beta n  + \theta  ) & 0 &  \\ 
       \vdots & 0 &\ddots  & 
    \end{pmatrix},
\end{equation}
it is known \cite{klein_anderson_2017} that all of the eigenstates of $H$ localize at sufficiently large $\lambda$ for almost every frequency $\beta$.
Conceptually, this property is similar to the way in which even infinitesimal disorder is sufficient to localize all eigenstates of a 1D model.

\subsection{Localization}

There are three different flavors of localization between which we distinguish. 
The first and simplest kind of localized state is the finitely supported\footnote{Such states are also sometimes called compact localized states.} eigenstate, which is clearly normalizable. A model may have an extensive number of finitely supported eigenstates at a particular energy, forming a flat band \cite{Vidal_1998,osborne2025gluedtreelatticescompact}. Often, such states arise because of delicate interference effects, and they are usually not robust to perturbations or interactions in the many-body case.  
A state $\psi$ is finitely supported if $\psi_n \psi_m = 0$
for $|n-m|$ sufficiently large. 

Owing to the sensitivity of finitely supported eigenstates to perturbations, it is necessary to consider a more permissive notion of localization. We say that a state $\psi$ is exponentially localized if there exists $\gamma >0$ such that 
\begin{equation}\label{eqn:exploc}
    |\psi_{n}| \leq e^{- \gamma |n|}
\end{equation}
for sufficiently large $|n|$. 
The largest $\gamma$ for which Eq.~\eqref{eqn:exploc} holds sets the scale on which localization occurs, and it is hence often called the inverse localization length. 
The quantity $\gamma$ is also sometimes colloquially termed a Lyapunov exponent, although $\gamma$ and the Lyapunov exponent are generally distinct objects, and in particular, a positive Lyapunov exponent does not automatically confer exponential localization.

A third possibility is that a state can be power-law localized with 
\begin{equation}
    |\psi_n| \leq C |\arg\max_{m}(|\psi_m|) - n|^{-\alpha}
\end{equation}
for some positive constants $C$ and $\alpha$, and for sufficiently large $|n|$. 
If $\alpha$\,$>$\,$1/2$, power-law localized states are normalizable. If $\alpha$\,$<$\,$1/2$, however, they need not be. 

One of the tasks that we will be concerned with is distinguishing between these cases numerically. 
To that end, we employ the inverse participation ratio, 
\begin{equation}
    \ipr(\psi) =\sum_{n}\left| \frac{\psi_n}{\left(\sum_m |\psi_m|^2\right)^{1/2}}\right|^4.
\end{equation}
For a state that fails to be normalizable, the denominators diverge, and accordingly, the $\ipr$ vanishes.
The scaling of the inverse participation ratio of an eigenstate with system size is a standard diagnostic for localization. 
Explicitly, as the system size $N$ varies, we define the fractal dimension for a state $\psi$ as the finite-size scaling exponent of the $\ipr$,
\begin{equation}
    \ipr(\psi) \sim N^{-D_2(\psi)}.
\end{equation}
A state with fractal dimension $D_2(\psi)$\,$=$\,$1$ is delocalized, while a state with $D_2(\psi)$\,$=$\,$0$ is localized (either exponentially or with a normalizable power law). If a state has fractal dimension somewhere between zero and one, we say that it is critical.
In practice, we can estimate $D_2(\psi)$ by diagonalizing finite-sized instances of some candidate model and calculating the $\ipr$ of eigenstates with similar energies. 
We remark that finitely supported eigenstates in a flat band confound the $\ipr$  and fractal dimension because numerical diagonalization generally chooses an arbitrary linear combination of degenerate states, which is almost always extended. In the models that we present, these flat-band states can be found exactly. 
Owing to this exact solvability of the finitely supported eigenstates, this will not be an obstacle.

\section{Single-particle localization}
In this section, we introduce and analyze several families of 1D hopping models subject to incommensurate magnetic fluxes. 

\subsection{Glued tree chain lattices}
Glued tree lattices \cite{osborne2025gluedtreelatticescompact} in 1D are very special; under the effects of a periodic flux, the entire spectrum can become finitely supported. In other words, all of the bands of the translation-invariant glued tree chain can be made exactly flat by external fields. 
The mechanism for this is that the lattices generally have ``choke'' points at which several paths destructively interfere exactly and completely.  

\begin{figure*}[t]
    \centering
    \includegraphics[width=\linewidth]{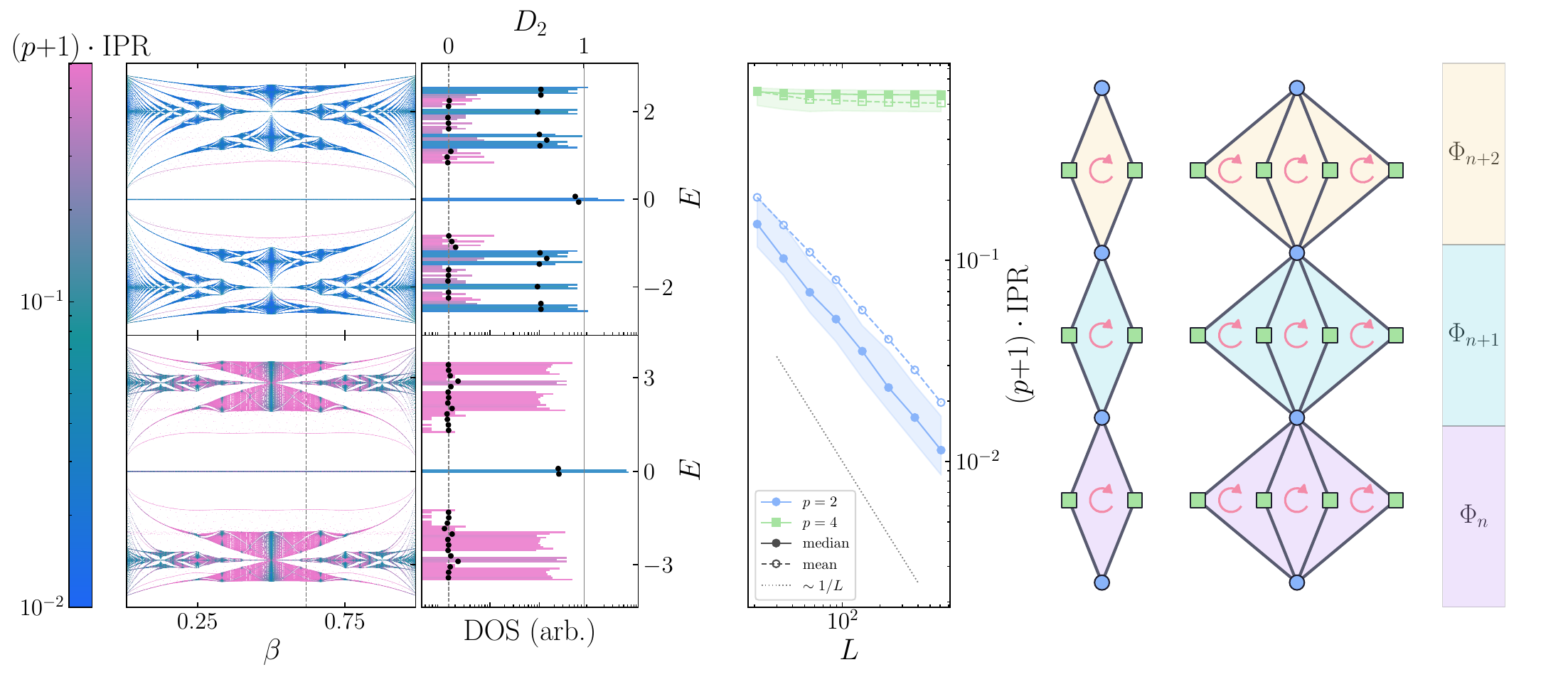}
    \caption{Spectral and localization properties of glued tree chains with quasiperiodic Peierls phases. Each unit cell consists of one hub site connected to $p$ branch sites.
{Left:} Hofstadter butterflies for $p=2$ (top) and $p=4$ (bottom), with eigenstates colored according to $(p+1)\cdot\mathrm{IPR}$ on a logarithmic scale, computed for $L=300$ unit cells. The dashed line marks $\beta=1/\varphi$, where $\varphi=(1+\sqrt{5})/2$. At this value, we additionally show the density of states, computed for $L=300$ and averaged over $20$ values of the phase $\theta$, together with the fractal dimension $D_2$, extracted from finite-size scaling of the mean IPR in each energy bin for system sizes $L\in\{89,\ldots,610\}$.
{Center:} Median (solid lines, filled markers) and mean (dashed lines, hollow markers) of $(p+1)\cdot\mathrm{IPR}$ as a function of system size $L$ at $\beta=1/\varphi$, excluding states with $|E|<0.1$. Shaded regions indicate the interquartile range, while the dotted line shows the $\sim 1/L$ scaling expected for extended states.
{Right:} Schematics of the lattices for $p=2$ and $p=4$. Colored faces indicate the quasiperiodic flux $\Phi_n=2\pi\beta n+\theta$, with arrows denoting the chosen sign convention.   
    }
    \label{fig:fig1}
\end{figure*}
Fix an integer $p > 1$ and define the $(p+1) \times (p+1)$ matrices
\begin{equation}\label{eqn:pchain}
    \begin{aligned}
        V_n(\theta) &= 
        \begin{pmatrix}
            0 & 1 & 1 & \dots  &1  \\ 
            1 & 0 & 0 & \dots  &0  \\ 
            1 & 0 & 0 & \dots  &0  \\
            \vdots &\vdots &\vdots &  &\vdots   \\
            1 & 0 & 0 & \dots & 0 
        \end{pmatrix}, \\ 
        T_n(\theta) &= 
        \begin{pmatrix}
            0 & 0 & \dots & 0 \\
            e^{\mathrm i \frac{(p-1)}{2} \Phi_n} & 0 & \dots & 0 \\
            e^{\mathrm i \frac{(p-3)}{2} \Phi_n} & 0 & \dots & 0 \\
            \vdots & \vdots & & \vdots \\
            e^{-\mathrm i \frac{(p-3)}{2} \Phi_n} & 0 & \dots & 0 \\
            e^{-\mathrm i \frac{(p-1)}{2} \Phi_n} & 0 & \dots & 0
        \end{pmatrix}
    \end{aligned}
\end{equation}
that specify the Hamiltonian as written in Eq.~\eqref{eqn:blocktr}. 
The $p = 2$ and $p = 4$ glued tree lattices are drawn in Fig.~\ref{fig:fig1}.

The $p=2$ glued tree chain has been studied in the context of quasicrystalline localization in a few settings. Recently, Ref.~\cite{kumar2026inapplicabilityavilastheorydiamond} has pointed out that the machinery of ``almost constant cocycles'' fails to detect mobility edges on the $p = 2$ glued tree lattice subject to quasiperiodic on-site potentials, while Refs.~\cite{Lee_2023,ahmed_flat-band-based_2022} have considered an on-site quasiperiodic potential in the presence of periodic magnetic fields, noting the pervasiveness of critical eigenstates (which we also identify, albeit without on-site potentials). 

We emphasize the presence of the extensive degeneracy at $E = 0$ for both the $p = 2$ and $p = 4$ cases. For the glued tree chain with arbitrary $p$, the degeneracy of the flat band at zero energy is $p-1$ per unit cell in the bulk, as argued in App.~\ref{app:flatband}. Furthermore, we can construct a strictly local basis for the set of states at zero energy, thereby demonstrating the coexistence of finitely supported flat-band eigenstates with states in the usual quasicrystalline trichotomy. 

Since glued tree lattices are bipartite, $H^2$ block-diagonalizes. 
In particular, one of the diagonal blocks of $H^2$ gives rise to the properly 1D hopping operator\footnote{The ratios of sines in Eq.~\eqref{eqn:1deff} are defined by continuity. For an open chain of $L$ unit cells, the diagonal entries of this block of $H^2$ at the two endpoints are $p$ rather than $2p$.}
\begin{equation}\label{eqn:1deff}
    (E^2 - 2p) \psi_n = \frac{\sin(p \Phi_n /2 )}{\sin(\Phi_n/2)} \psi_{n+1} +  \frac{\sin(p \Phi_{n-1} /2 )}{\sin(\Phi_{n-1}/2)} \psi_{n-1}.
\end{equation}
If an eigenstate of Eq.~\eqref{eqn:1deff} is localized at $\mu = (E^2 - 2p)$, then there must be a pair of states at
$E = \pm \sqrt{\mu + 2p}$ of Eq.~\eqref{eqn:blocktr} that are also localized, owing to bipartiteness. The only possible exceptions are states with no support on the sites of degree $2p$, which we have shown to be finitely supported and which exist at zero energy.

With $p = 2$, Eq.~\eqref{eqn:1deff} reduces to 
\begin{equation}\label{eqn:p2}
    \frac{1}{2}(E^2 - 4) \psi_n = \cos(\Phi_n/2) \psi_{n+1} + \cos(\Phi_{n-1}/2)\psi_{n-1},
\end{equation}
which is the off-diagonal Aubry-Andr\'e-Harper model \cite{PhysRevLett.109.106402}. 
This model is known to be critical in the entirety of its spectrum by virtue of the fact that it maps unitarily onto the AAH model at the critical point \cite{jitomirskaya2019criticalmathieuoperatorhidden}, with the off-diagonal frequency $\beta/2$ corresponding to the AAH frequency $\beta$.
A further result of Ref.~\cite{jitomirskaya2019criticalmathieuoperatorhidden} is that the Hausdorff dimension of the spectrum of Eq.~\eqref{eqn:p2} is bounded above by $1/2$, which forbids an absolutely continuous spectrum. 
For $p>2$, however, the theorems used to demonstrate the criticality of Eq.~\eqref{eqn:p2} are not applicable.

Singular Jacobi operators of this kind can, in principle, be treated with the methods of Ref.~\cite{jitomirskaya_analytic_2011}. However, the complexification arguments that furnish lower bounds on the Lyapunov exponent (reviewed in App.~\ref{app:complex}) require that the hopping coefficients be nonvanishing on some complex strip of finite width about the real line, whereas Eq.~\eqref{eqn:1deff} has hopping coefficients with several zeros along the real line (see App.~\ref{app:pg3}).
Here, we use numerical techniques to argue that $p = 3$, like $p = 2$, is critical in its entire spectrum and that, for $p > 3$, all eigenstates of Eq.~\eqref{eqn:1deff} are exponentially localized, with the exception of those at $E = \pm \sqrt{2 p}$.
Figure~\ref{fig:fig1} shows numerical evidence that eigenstates at nonzero energies exponentially localize for $p = 4$ glued tree lattices,
with further evidence presented in App.~\ref{app:pg3}.

In this family of models, the integer $p$ plays a role comparable to the on-site potential strength $\lambda$ from the AAH model. For sufficiently large $p$, eigenstates are apparently Anderson localized. Unlike in the case of the AAH model, there is no sense in which the parameter mediating the localization transition can be smoothly deformed. 

If $\beta$ is rational and the orbit contains $\Phi_m=2\pi r/p$ modulo $2\pi$ for some $m\in\mathbb Z$ and $r\in\{1,\ldots,p-1\}$, then the spectrum of the Hamiltonian consists entirely of a finite set of energies: instead of delocalization with continuous dispersive bands, the spectrum becomes fully localized, but the mechanism of localization is fine-tuned Aharonov-Bohm caging rather than the exponential localization characteristic of Anderson physics.

\subsection{Exact mobility edge}
It is natural to ask whether it is possible to use existing theorems and techniques to conclusively show exponential localization on any lattice at all due to incommensurately periodic magnetic fields. 
The concern is primarily that models with several sites in a unit cell generally have a hierarchy of Lyapunov exponents, all of which must be positive in order to show that all of the eigenstates localize~\cite{klein_anderson_2017}.
Bounds arising from Avila's global theory and generalizations thereof \cite{avila_global_2009,marx_dominated_2014,jitomirskaya_analytic_2011,Avila_2014,Ge_Jitomirskaya_You_Zhou_2026} have been incompletely extended to cover such cases.
Furthermore, the scalar reductions of the models arising in the previous subsection are actually singular Jacobi operators rather than Schr\"odinger operators, which complicates the issue.

Here, we present an easily extensible family of models where Avila's global theory provides enough leverage to show an exact mobility edge. 

\begin{figure}
    \centering
    \includegraphics[width=\linewidth]{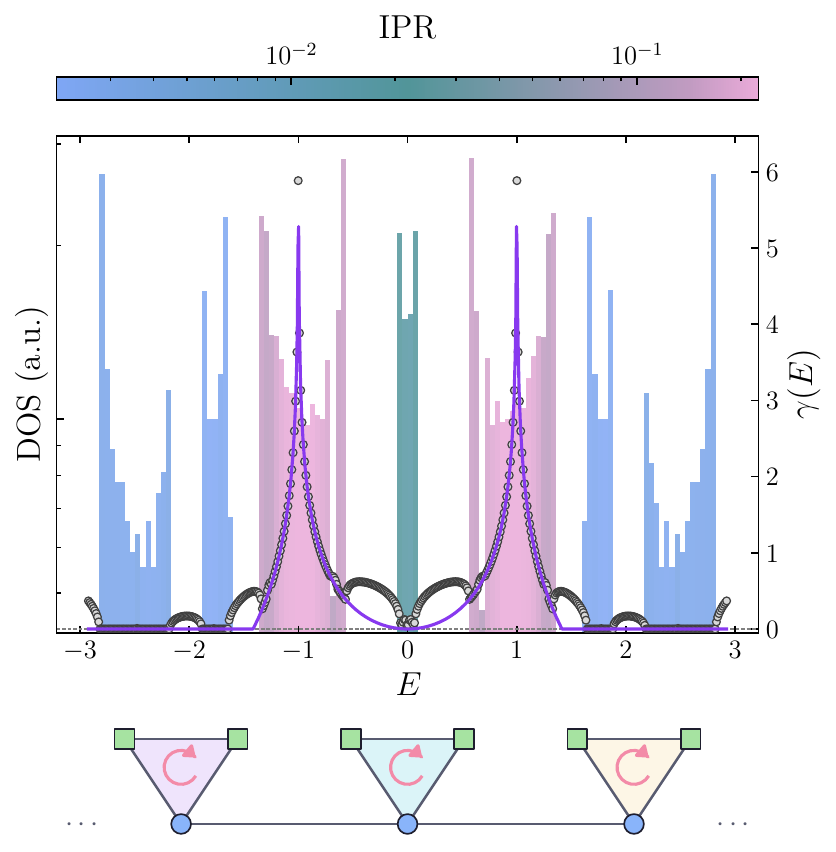}
    \caption{Top: Density-of-states (DOS) histogram at $\beta = 1/\varphi$ averaged over 20 phase offsets $\theta$ at $L = 300$ unit cells ($N = 900$ sites), with bars colored by mean IPR. 
    The Lyapunov exponent $\gamma(E)$ is shown on the right axis: black points are computed numerically via the transfer matrix over $5 \times 10^5$ unit cells, and the purple curve is the analytical result in Eq.~\eqref{eqn:lyapexact}, continued through spectral gaps for comparison. 
    Bottom: Lattice schematic with colored faces indicating the quasiperiodic flux $\Phi_n = 2\pi\beta n + \theta$ threading each plaquette. 
    }
    \label{fig:fig2}
\end{figure}

The Hamiltonian pictured in Fig.~\ref{fig:fig2}
has 
\begin{equation}\label{eqn:pstub}
\begin{aligned}
    T_n(\theta) &= \begin{pmatrix}
        1 & 0 & 0  \\ 
        0 & 0 & 0  \\ 
        0 & 0 & 0  \\ 
    \end{pmatrix}, \\ 
    V_n(\theta) &= \begin{pmatrix}
        0 & e^{\mathrm i \Phi_n/2} & e^{-\mathrm i \Phi_n/2} \\ 
        e^{-\mathrm i \Phi_n/2}  & 0 & 1 \\
        e^{\mathrm i \Phi_n/2} & 1 & 0
    \end{pmatrix}.
    \end{aligned}
\end{equation}
In contrast to the model in Eq.~\eqref{eqn:pchain}, the model given by Eq.~\eqref{eqn:pstub} cannot be reduced to a 1D chain by taking a finite power of the Hamiltonian matrix. 
Instead, we are required to invoke the Schur complement. 
Integrating out the sites in the triangular pendants, for $E\neq\pm1$, yields the 1D effective energy-dependent hopping operator
\begin{equation}\label{eqn:schrodeff}
   E \psi_n =  \psi_{n+1} + \psi_{n-1} + V^{\text{eff}}_n(E) \psi_n,
\end{equation}
with 
\begin{equation}
    V^{\text{eff}}_n(E) = \frac{2 E + 2 \cos(\Phi_n)}{E^2 -1}.
\end{equation}
An application of Avila's global theory \cite{avila_global_2009}, which we review in App.~\ref{app:complex}, immediately reveals an energy-dependent Lyapunov exponent:
\begin{equation}\label{eqn:lyapexact}
    \gamma(E) = \max(0, -\log|E^2 - 1|),
\end{equation}
which we have plotted alongside numerically calculated Lyapunov exponents in Fig.~\ref{fig:fig2}.

Notably, Eq.~\eqref{eqn:lyapexact} is exact: the effective Schr\"odinger operator in Eq.~\eqref{eqn:schrodeff} is simply the AAH model with an energy-dependent on-site potential. 
In the context of operator cocycle complexification arguments, this energy dependence lies on equal footing with ordinary parameter dependence, and hence poses no obstruction. 
We remind the reader that Eq.~\eqref{eqn:lyapexact} is only correct for energies actually in the spectrum of the Hamiltonian, and energies outside of the spectrum have a positive Lyapunov exponent even for the ordinary 1D chain.

The mobility edge can now be read off directly from Eq.~\eqref{eqn:lyapexact}. The effective AAH potential is of strength $\lambda = 2/(E^2-1)$, whose magnitude exceeds the critical value $|\lambda|=2$ for $0<|E|<\sqrt{2}$ and falls below it for $|E|>\sqrt{2}$. Eigenstates with $0<|E|<\sqrt{2}$ are therefore exponentially localized with localization length $1/\gamma(E)$, while those with $|E|>\sqrt{2}$ are extended, and the two are separated by exact mobility edges at $E = \pm\sqrt{2}$ (see App.~\ref{app:3stub}). This is borne out by the $\ipr$ of the eigenstates in Fig.~\ref{fig:fig2}. At $\beta = 1/\varphi$, the mobility edges themselves happen to fall in spectral gaps, so that each band is either entirely localized or entirely extended.

It is straightforward to formulate generalizations of this model that also become Schr\"odinger operators with energy-dependent quasiperiodic effective on-site potentials under the Schur complement. For instance, one could increase the number of edges in the pendant loop (here, a triangle) or, alternatively, allow the magnitude of the hopping coefficients among edges in the pendants to differ from the hopping magnitudes along the spine. This set of extensions readily gives rise to an infinite family of magnetic quasicrystals with exactly tractable mobility edges (see App.~\ref{app:3stub}).

\subsection{Typical behavior}

Generic magnetic quasicrystals fall outside the scope of what can be exactly described by existing theorems. 
For 1D quasiperiodic Schr\"odinger operators, the absence of criticality is prevalent in the measure-theoretic sense \cite{avila_global_2009}, and it is not obvious that the same statement holds in the present case.
Apart from the genuine 1D chain, which trivially cannot localize under quasiperiodic Peierls phases, it is nontrivial to construct a model that is entirely extended in its spectrum. 
As a pair of immediate examples, the square and triangular ladder lattices in App.~\ref{sec:ladder} come to mind. It so happens that the square ladder lattice has an apparently extensive set of critical wave functions. 
In contrast to the AAH model, we find that magnetic quasicrystals can have localization properties that are dependent upon the particular frequency $\beta$. 
An example of this is the triangular ladder lattice, which has mostly extended eigenstates for most values of $\beta$ but an appreciable number of genuinely localized eigenstates for some $\beta$. 

\begin{figure*}[t]
 \centering
 \includegraphics[width=\linewidth]{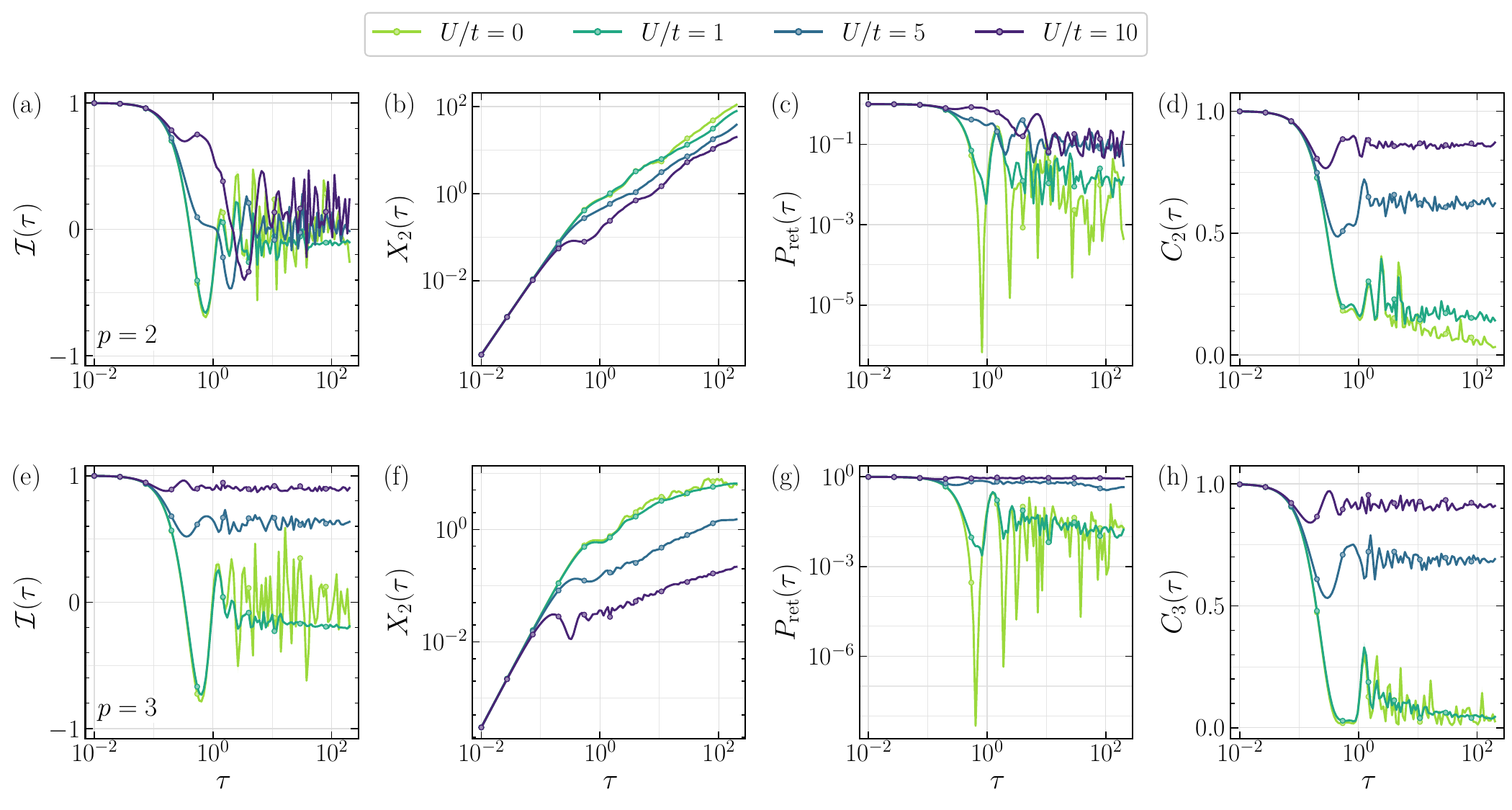}
 \caption{Bosonic dynamics in the few-body regime of a glued tree chain lattice with open boundaries at $\beta=1/\varphi$. Top: $p=N_b=2$, $L=64$, $N=193$. Bottom: $p=N_b=3$, $L=12$, $N=49$. The columns correspond to the different observables in Eq.~\eqref{eqn:dynobs}. Each curve represents the mean over $100$ independent uniform phases $\theta\in[0,2\pi)$ for every $U/t$.}
 \label{fig:interacting}
\end{figure*}

\section{Interaction-induced delocalization}\label{sec:interactions}
The localization phenomena identified in the models above are purely interference effects, which are known to be fragile against interactions \cite{vidal_interaction_2000}. Numerical tests of the \textit{statics} and spectra in the interacting regime, however, run into difficulties. Specifically, the few-body Hilbert space is statistically dominated by localized modes, some of which remain exact zero-energy eigenstates for every $U$, and the resulting degeneracies spoil the level statistics (as described in App.~\ref{app:fewbody}). We therefore switch viewpoints and study the \textit{dynamics} of $N_b=p$ bosons on open glued tree chains with an even number $L$ of unit cells and $N=(p+1)L+1$ sites  using exact diagonalization. Our Hamiltonian is given by
\begin{equation}\label{eqn:interacting}
 \widehat H=t\sum_{x,y=0}^{N-1}H_{xy}b_x^\dagger b^{\pdagger}_y
 +\frac U2\sum_x\hat n^{\pdagger}_x(\hat n^{\pdagger}_x-1),
\end{equation}
where $H$ is the unit-hopping matrix of Eqs.~\eqref{eqn:blocktr} and \eqref{eqn:pchain}, $b_x$ annihilates a boson on site $x$, $t$\,$>$\,$0$ is the hopping scale, $\hat n_x$\,$\equiv $\,$b_x^\dagger b_x$, and $U$\,$\geq$\,$0$ is the on-site repulsion. The bosons start clustered on the central hub $x_0=(p+1)L/2$, in the initial state $\lvert\psi_0\rangle=\lvert N_b:x_0\rangle$ with $\lvert N_b:x\rangle \equiv (b_x^\dagger)^{N_b}\lvert0\rangle/\sqrt{N_b!}$, which has no overlap with the flat band (see App.~\ref{app:flatband}).

Calculating $\lvert\psi(\tau)\rangle=e^{-\mathrm i\widehat H\tau/t}\lvert\psi_0\rangle$ in the fixed-$N_b$ sector (where we introduce  $\tau=t\,t_{\mathrm{phys}}/\hbar$ as our unit of time), we evaluate and track several quantities:
\begin{equation}\label{eqn:dynobs}
\begin{aligned}
 \mathcal I(\tau)&=\frac1{N_b}\sum_x s_x\langle\hat n_x\rangle_\tau,\\
 X_2(\tau)&=\frac1{N_b}\sum_x[n(x)-L/2]^2\langle\hat n_x\rangle_\tau,\\
 P_{\mathrm{ret}}(\tau)&=|\langle\psi_0|\psi(\tau)\rangle|^2,\\
 C_{N_b}(\tau)&=\sum_x|\langle N_b:x|\psi(\tau)\rangle|^2,
\end{aligned}
\end{equation}
where $s_x$\,$=$\,$+1$ on hubs and $-1$ on branches, and $n(x)=\lfloor x/(p+1)\rfloor$ is the unit cell of site $x$. 
Intuitively, the imbalance $\mathcal{I}$ measures how the density is shared between the two sublattices, $X_2$ is the second moment of the density about the initial unit cell, and the cluster weight $C_{N_b}$ is the probability that all bosons occupy one site. Subtracting the return probability $P_{\mathrm{ret}}$ from $C_{N_b}$ thus gives the probability of finding such an on-site cluster elsewhere, away from $x_0$.

Figure~\ref{fig:interacting} showcases the resulting dynamics at $\beta = 1/\varphi$. In the noninteracting limit at $U=0$, the bosons spread independently through the critical states. As a result, $X_2$ grows steadily, limited only by the length of the chain, and $\mathcal I$ oscillates about zero. Since every single-particle eigenstate at nonzero energy has equal weight on the two sublattices (by virtue of bipartiteness), the dispersive states contribute to $\mathcal I$ only through coherences between partner states at $\pm E$. Once these have dephased, a negative mean $\mathcal I$ requires weight in the flat band, which lives on the branches alone. This is precisely what we find at $U/t=1$. In this case, the oscillations of $\mathcal{I}$ are damped, and it relaxes to $\approx-0.1$ for $p=2$ and $\approx-0.2$ for $p=3$, while $X_2$ grows at a rate comparable to that of free particles. Interactions therefore scatter weight from the dispersive states into the finitely supported states of the flat band and, the coupling being Hermitian, out of them as well. In other words, the flat band is no longer an inert localized sector. 

For strong repulsions, the picture changes considerably because the bosons bind. To wit, at $U/t=10$ and late times ($\tau$\,$=$\,$ 200$), we see that the probability of all bosons sharing a site seemingly saturates to $C_2\simeq0.871$ and $C_3\simeq0.910$.
Such a cluster can move only through a process of order $N_b$ in the hopping, with amplitude $t_{\mathrm{cl}}=N_b t^{N_b}/[(N_b-1)!\,U^{N_b-1}]$, and it carries $N_b$ times the Peierls phase of a single boson. For the doublon, $t_{\mathrm{cl}}=2t^2/U = 0.2 t$, resulting in a slower clock, wherefore $X_2$ qualitatively follows the noninteracting curve at a reduced rate. The bound pair spreads through the chain as a composite particle and is found away from $x_0$ with probability $0.666\pm0.008$ (here and below, uncertainties are standard errors over phases). In contrast, from the $p = 3$ data at strong repulsion, we notice that the triplon barely moves over the same duration (the corresponding probability is only $0.037\pm0.003$); concomitantly, $X_2$ remains well below unity, and $\mathcal{I}$ stays close to its initial value. Part of the explanation is, of course, a separation of time scales, since the triplon hops with $t_{\mathrm{cl}} = 3t^3/(2U^2) = 0.015t$, whereas the second-order energy shift of a cluster, $N_b\,z\,t^2/[(N_b -1)U]$ on a site of degree $z$, differs between hubs ($z = 2p$) and branches ($z = 2$) by $6t^2/U = 0.6t$. Transfer from one hub to the next is then off-resonant and proceeds at a rate of order $t_{\mathrm{cl}}^2 /(0.6t) \sim 4 \times 10^{-4} t$, which moves only a few percent of the weight for $\tau \le 200$, in line with the probability quoted above (the corresponding detuning for the doublon, $4t^2/U = 0.4t$, is comparable to $t_{\mathrm{cl}}$). It is unlikely, though, that this is the whole story. In fact, the small $X_2$ continues to grow with both time and system size (see App.~\ref{app:fewbody}), although much of it appears to be carried by the weight outside the on-site cluster ($1-C_3\approx0.09$), and the present data do not tell us whether the triplon eventually spreads as the doublon does.

Taken together, these results lead us to conclude that, for both $p=2$ and $p=3$, interactions act against localization, in line with the ordinary AAH model \cite{Doggen_2019}, despite the fact that, here, there is no potential for them to compete with. A few remarks are in order. We caution first that the hub state is not localized at $U/t=0$ and that repulsion does not necessarily make it spread any faster; rather, what interactions remove is the protection of the flat band. Our calculations are also restricted to a fixed number of particles and to finite times, so they do not address MBL at finite density. Finally, given the severe finite-size limitations of full diagonalization, we have not treated $p > 3$, where the dispersive single-particle states are themselves exponentially localized.
It is unclear how large $p$ must be for localization to survive interactions, but we expect that it eventually does, as it does in the AAH model for strong potentials \cite{Mastropietro_2015}.

\section{Discussion}

The interplay of periodic microscopic lattice topology with a linearly increasing magnetic field in one dimension gives rise to a quasiperiodicity that can mediate localization transitions analogous to the AAH metal-insulator transition. 
In this work, we have presented a number of examples of such magnetic quasicrystals. In doing so, we have introduced models with exact mobility edges in the absence of on-site potentials, and models with nearly every eigenstate exponentially localized. 
We also find that the flat bands of the underlying lattice can persist under the effects of an external field, leading to the coexistence of finitely supported and exponentially localized or critically localized states.
In keeping with conventional wisdom, we observe that interactions tend to mediate delocalization in magnetic quasicrystals. Quite generically, on-site repulsion mixes the flat band with the dispersive states and does not arrest the spreading of two bosons at any strength that we have studied.

The complex and varied phenomenology that we uncover here is ripe for future work in both pure mathematics and physics. For instance, 
to the best of our knowledge, there does not exist a systematic formalism whereby the Lyapunov exponents of Eq.~\eqref{eqn:1deff} for large $p$ can be shown to be positive on the spectrum. It would be interesting to know whether modern arguments relying on Aubry duality \cite{Ge_Jitomirskaya_You_Zhou_2026} could be generalized to the singular Jacobi case or whether the problem of real roots in hopping coefficients poses an insurmountable obstruction.

\begin{acknowledgments}
 The authors would like to thank Ciro Salcedo and R. N. Bhatt for helpful discussions. 
 We acknowledge support from the Co-design Center for Quantum Advantage ($\text{C}^2\text{QA}$) under Contract No. DE-SC0012704.

 Princeton University Professor Andrew Houck is also a consultant for Quantum Circuits Incorporated (QCI). Due to his income from QCI, Princeton University has a management plan in place to mitigate a potential conflict of interest that could affect the design, conduct and reporting of this research.
\end{acknowledgments}

\begin{appendix}

   \section{Flat bands in the glued tree lattices under incommensurate magnetic fields}\label{app:flatband}
   Glued tree chains are bipartite, admitting a Hamiltonian matrix decomposition of the form
   \begin{equation}
       H = \begin{pmatrix}
           0 & X \\ 
           X^\dagger & 0
       \end{pmatrix}.
   \end{equation}
   As discussed in the main text, one part of the bipartition consists of the ``hub'' vertices of degree $2p$. The remaining vertices are $p$ in number per unit cell, but $X^\dagger X$ and $X X^\dagger $ must have the same rank, so there are $p - 1$ null vectors per unit cell (more precisely, $(p-1)L-1$ on an open chain of $L$ unit cells). All that remains is to show that the set of null vectors of $H$ admits a basis of finitely supported states. 

   Let the sites labeled by $(p+1) n$, for integer $n$, be the hub vertices of degree $2p$. Among the sites $(p+1)n + 1, (p+1)n + 2 , \dots, (p+1)n + p$, there are $p$ degrees of freedom. Since each of these sites is connected to site $(p+1)n$ and site $(p+1)(n+1)$, there are two constraints that must be satisfied for 
   \begin{equation}
      |\psi\rangle = \sum_{i = (p+1) n+1}^{(p+1)(n+1) - 1} \psi_i |i \rangle
   \end{equation}
   to be a null vector of $H$, and hence there are $p-2$ linearly independent null vectors living on these $p$ sites alone. 
   To conclude our construction of the basis of null vectors, observe that the sites in the $(n+1)$th unit cell also contribute $p -2$ null vectors strictly within the $(n+1)$th unit cell. However, there are $2p$ degrees of freedom among the degree-$2$ vertices of the two cells, and only three constraints must be satisfied by null vectors, giving $2p-3$ null vectors, one more than the two single-cell subspaces supply. Inducting in this way from left to right produces $(p-2)L+(L-1)=(p-1)L-1$ independent null vectors on an open chain of $L$ unit cells, which is sufficient to construct a basis of null vectors.

   Note that this sort of combinatorial argument may also be used to show that $E = 0$ is not an eigenvalue of the hub block $XX^\dagger$ of $H^2$ underlying Eq.~\eqref{eqn:1deff}, though it may still be in the spectrum. This is because, if there were an eigenvector $u$, then $X^\dagger u=0$, and the matrix 
   \begin{equation}
   M = 
   \begin{pmatrix}
       1 & 1 & \dots & 1 \\ 
       1 & e^{\mathrm i \Phi_n} & \dots & e^{\mathrm i \Phi_n (p-1)}
       \end{pmatrix}
   \end{equation}
   would have to have rank $1$ in every unit cell, which is clearly false for irrational $\beta$.

   \section{Lyapunov exponents by cocycle complexification}\label{app:complex}
   A recurrence relation of the form 
   \begin{equation}
       E \psi_n = t_n(\theta) \psi_{n+1} + t_{n-1}(\theta) \psi_{n-1} + V_n(\theta) \psi_n
   \end{equation}
   can be reorganized according to its transfer matrices: 
   \begin{equation}
       \begin{pmatrix}
           \psi_{n+1} \\ 
           \psi_n
       \end{pmatrix}
       =
       \begin{pmatrix}
           \frac{E - V_n(\theta)}{t_{n}(\theta)} & - \frac{t_{n-1}(\theta)}{t_n(\theta)} \\ 
           1 & 0 
       \end{pmatrix}
       \begin{pmatrix}
           \psi_{n} \\ 
           \psi_{n-1}
       \end{pmatrix},
   \end{equation}
   and eigenstates can thus be calculated iteratively. 
   We write
   \begin{equation}
       \mathcal{T}_n(\theta) \equiv  
       \begin{pmatrix}
           \frac{E - V_n(\theta)}{t_{n}(\theta)} & - \frac{t_{n-1}(\theta)}{t_n(\theta)} \\ 
           1 & 0 
       \end{pmatrix}.
   \end{equation}
   The Lyapunov exponent of this operator, $\gamma(E)$, is defined as
   \begin{equation}\label{eqn:gammadefn}
   \begin{aligned}
       \gamma(E) &=\hspace*{-0.1cm} \lim_{N\rightarrow \infty} \frac{1}{P N}\int_0^P \hspace*{-0.25cm} \mathrm d\theta \log\|
        \mathcal{T}_{N}(\theta)
        \mathcal{T}_{N-1}(\theta)
        \dots 
        \mathcal{T}_{1}(\theta)
       \|.
   \end{aligned}
   \end{equation}
   Here, $P$ is a phase period of the coefficients, and $\|\cdot\|$ is the Euclidean operator norm.
   The positivity of the Lyapunov exponent does not automatically imply localization in general, but it does sometimes imply exponential localization of eigenstates.
   On the other hand, a positive Lyapunov exponent is required for exponential localization.

   Define the phase-averaged complexified exponent
   \begin{equation}
   \begin{aligned}
       L(E,\epsilon) &= \lim_{N\rightarrow \infty} \frac{1}{P N}\int_0^P\mathrm d\theta\\
       &\quad\log\|
        \mathcal{T}_{N}(\theta+\mathrm i\epsilon)
        \mathcal{T}_{N-1}(\theta+\mathrm i\epsilon)
        \dots
        \mathcal{T}_{1}(\theta+\mathrm i\epsilon)
       \|.
   \end{aligned}
   \end{equation}
   In practice, it is possible to bound $\gamma(E)=L(E,0)$ from below using the dependence of $L(E,\epsilon)$ on $\epsilon$. To this end, the rich and storied literature around the AAH operator is very helpful. 
   In Refs.~\cite{avila_global_2009,jitomirskaya_analytic_2011,Avila_2014}, it was shown that $L(E,\epsilon)$ is piecewise linear, continuous, and convex in $\epsilon$ for irrational frequency, provided that the functions $t_n(\theta)$ and $V_n(\theta)$ are analytic and $t_n(\theta)$ is bounded away from zero in a complex strip of finite width about the real line. 
   Hence, a program for analytically bounding the Lyapunov exponent of a 1D hopping operator is to
   first calculate 
   \begin{equation}\label{eqn:disast}
      \mathcal T_n(\theta + \mathrm i \epsilon) = \tilde{\mathcal{T}}_n(\theta, \epsilon) + o\big(\|\tilde{\mathcal{T}}_n(\theta,\epsilon)\|\big)
   \end{equation}
   in the limit of large $\epsilon$, and then use the resulting simplifications to produce 
   \begin{equation}\label{eqn:goal}
    L(E,\epsilon) = \delta + Z |\epsilon|,
   \end{equation}
   where $Z$ is a nonnegative integer. By convexity, we then have that 
   \begin{equation}\label{eqn:bound}
       \gamma(E) \geq \delta.
   \end{equation}
   
    In the case of Schr\"odinger operators, where $t_n(\theta) = 1$ and where $V_n(\theta)$ is bounded above, Eq.~\eqref{eqn:bound} is often tight.
    If, on the other hand, $t_n(\theta)$ is nontrivial, some caveats apply. 
    The continuity, piecewise linearity, and convexity of $L(E,\epsilon)$ are not immediate when $t_n(\theta)$ is allowed to vanish or when $V_n(\theta)$ is unbounded. 
    It may also happen in generic models that Eq.~\eqref{eqn:bound} fails to be tight, as we will argue is the case for Eq.~\eqref{eqn:1deff} in App.~\ref{app:pg3}.

   \section{Evidence for localization in the $p > 3$ glued tree lattices}\label{app:pg3}
   \begin{figure*}
     \includegraphics[width=\linewidth]{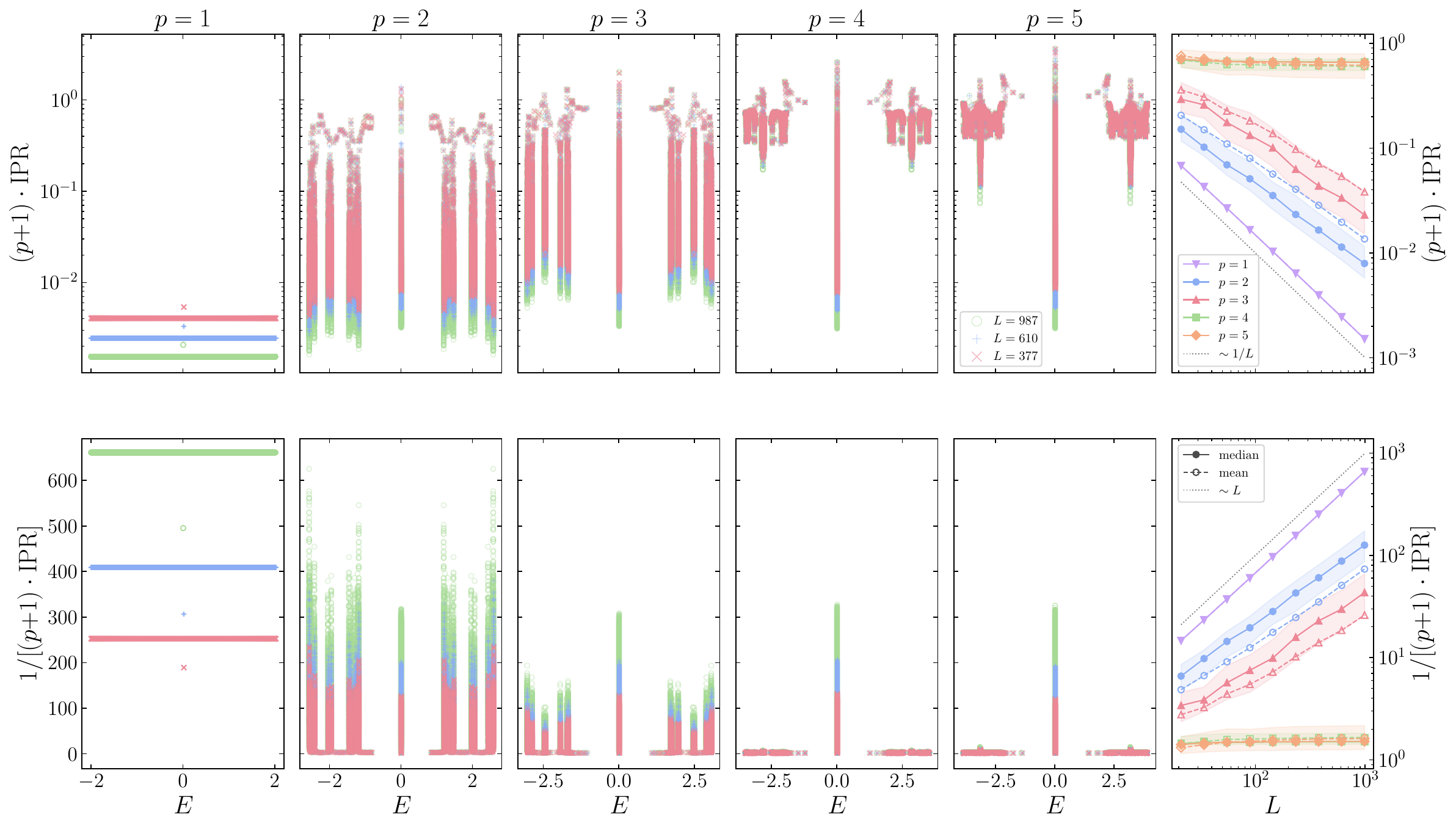}
     \caption{Scaling of the inverse participation ratio with system size as a function of energy. Top and bottom rows contain the same data visualized differently.  
     In the rightmost panels, solid lines and filled markers denote the median, dashed lines and hollow markers denote the mean, and the shaded region is the interquartile range. 
     Data is produced at  $\beta = 1/\varphi$ for $L \in \{21, 34, \dots, 987\}$ unit cells ($N = (p+1)L + 1$ sites) and averaged over 20 values of $\theta$. The rightmost panels exclude the flat band via $|E| > 0.1$.
     }\label{fig:iprscale-extended}
   \end{figure*}

    $\ipr$ statistics do not necessarily capture all of the localization properties of models. In general, localization properties are energy dependent, and averages of the $\ipr$ can be blind to, for example, a small but extensive set of states delocalizing while all others are exponentially localized. 
    Thus, it can be instructive to see the energy-resolved $\ipr$ scaling of all eigenstates. In Fig.~\ref{fig:iprscale-extended}, we have shown the energy-resolved $\ipr$ and its inverse for several values of $p$ alongside an extended version of the $\ipr$ summary statistics plot that appears in the main text. 

    In Fig.~\ref{fig:iprscale-extended}, $p = 1$ is simply the 1D chain, which is genuinely periodic. We include it as a point of contrast; $p = 1$ is extended in every eigenstate. 
    We see a qualitative change in behavior between $p=1$ and $p = 2$, which is known to be critical in its entire spectrum. 
    There is yet another qualitative change at $p = 4$, where, apart from $E = 0$, $\ipr$ values do not discernibly scale with size in any way. 
    The randomly chosen superposition of strictly local null vectors does, unsurprisingly, scale with system size.

\begin{figure*}
       \includegraphics[width=\linewidth]{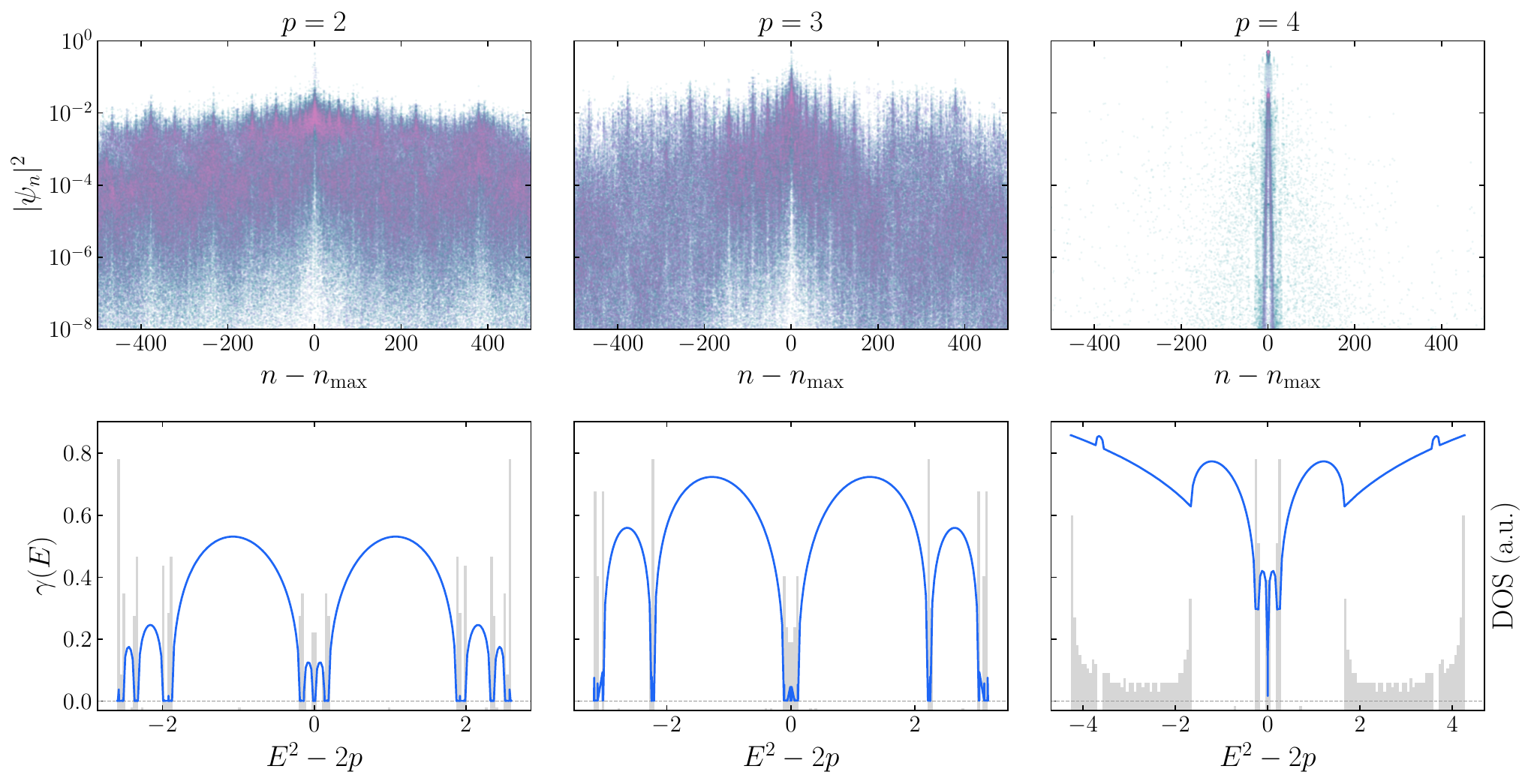}
       \caption{Eigenstate localization and numerical calculation of Lyapunov exponents for Eq.~\eqref{eqn:1deff}. Eigenstates are calculated with $L = 1000$, and Lyapunov exponents are calculated with $N = 10^5$ transfer steps. In regions where the density of states is nonvanishing, the Lyapunov exponent is calculated using the energy from Hamiltonian diagonalization.}\label{fig:state-and-lyap}
   \end{figure*}

   As we argued in the main text, the glued tree chain magnetic quasicrystal is localized in its spectrum if and only if Eq.~\eqref{eqn:1deff} is localized. To that end, we turn our attention toward Eq.~\eqref{eqn:1deff} explicitly. 
   In Fig.~\ref{fig:state-and-lyap}, we have overlaid the probability densities of eigenstates of Eq.~\eqref{eqn:1deff}, showing a sharp distinction between the $p =2$ critical behavior and the $p = 4$ exponentially localized behavior. 

   Unlike in the case of the AAH model, Eq.~\eqref{eqn:1deff} is allowed to have an energy-dependent Lyapunov exponent. The transfer matrix of Eq.~\eqref{eqn:1deff} is given by 
   \begin{equation}
       \mathcal T_n(\theta) = 
           \frac{1}{t_n(\theta)}
       \begin{pmatrix} 
       E^2 - 2p & - t_{n-1}(\theta) \\ 
       t_n(\theta) & 0 
       \end{pmatrix},
   \end{equation}
   with 
   \begin{equation}
       t_n(\theta) = \frac{\sin(p\Phi_n/2)}{\sin(\Phi_n/2)}.
   \end{equation}
   When $E= \pm \sqrt{2 p}$, 
   \begin{equation}
   \begin{aligned}
       \mathcal T_{n+1}(\theta)\mathcal T_n(\theta)
       &= -\frac{1}{t_{n}(\theta) t_{n+1}(\theta)}\\
       &\quad\times\begin{pmatrix}
           t_n(\theta)^2 & 0 \\
           0 & t_{n+1}(\theta) t_{n-1}(\theta)
       \end{pmatrix}.
   \end{aligned}
   \end{equation}
   By ergodicity, for irrational $\beta$ and almost every $\theta$, the even and odd subsequences ($r=0,1$) obey
   \begin{equation}
   \begin{aligned}
       &\lim_{N\rightarrow\infty}\frac1N\sum_{n=0}^{N-1}\log|t_{2n+r}(\theta)|\\
       &\quad=\frac1{2\pi}\int_0^{2\pi}\mathrm d\Phi\,
       \log\left|\frac{\sin(p\Phi/2)}{\sin(\Phi/2)}\right|.
   \end{aligned}
   \end{equation}
   An elementary change of variables provides 
   \begin{equation}
       \int_0^{2 \pi}\mathrm d\Phi\log|\sin(p\Phi/2)| = \int_0^{2 \pi}\mathrm d\Phi\log|\sin(\Phi/2)|,
   \end{equation}
   so both parity averages vanish. The logarithms of the two diagonal entries in the even-length transfer product are differences of these subsequence sums, and hence
   \begin{equation}
       \gamma(E=\pm\sqrt{2p})=0.
   \end{equation}
   We see that the recurrence relation corresponding to $E = \pm \sqrt{2 p}$ generally admits a pair of solutions in the infinite case, but does not generally admit any solutions in the finite-size case, where the two boundary conditions overdetermine it.

   The lower row of panels in Fig.~\ref{fig:state-and-lyap} shows 
   transfer matrix Lyapunov exponents for Eq.~\eqref{eqn:1deff} for a few values of $p$. On the spectrum of the $p = 2,3$ models, we see the vanishing of the Lyapunov exponent. 
   For $p = 4$, the localization length of eigenstates decreases with increasing $|E^2 - 2p|$. Accordingly, eigenstates with energies near $\pm\sqrt{2p}$ have larger localization lengths (although they have support on $\sim 10$ sites and are unambiguously localized), which leads to the apparent fringe in the top rightmost panel of Fig.~\ref{fig:state-and-lyap}. 
   Empirically, the calculated Lyapunov exponent for $p = 4$ fits exceedingly well to 
   \begin{equation}
       \tilde{\gamma}(\mu) = \log\log(2 |\mu| + c),
   \end{equation}
   with $\mu = E^2 - 2 p$ for $c \sim 3.8$. 
   Interestingly, the only points along the spectrum where $\tilde{\gamma}$ gives rise to an appreciable departure from the calculated data are near $\mu = 0$. 
   Since we have no formal way of extracting a Lyapunov exponent, we can only speculate as to the cause of this very strong agreement. The departure near $\mu = 0$ is unavoidable, however: the Lyapunov exponent remains continuous in $E$ even for singular cocycles of this kind \cite{jitomirskaya_analytic_2011} and vanishes at $\mu=0$, whereas $\log\log c>0$.

Broadly speaking, the inapplicability of Avila's global theory and generalizations thereof to these models is not of the sort where the method provides apparently correct results but nonetheless lacks a formal proof that it must always be so.
    The transfer matrices are singular at the real zeros of the hopping coefficients, but the regularized matrices $A_n(\theta)=t_n(\theta)\mathcal T_n(\theta)$ are analytic, and the phase-averaged Lyapunov exponent $L_A(E,\epsilon)$ of their products is continuous, piecewise linear, and convex in $\epsilon$ \cite{jitomirskaya_analytic_2011}. The two exponents are related by
    \begin{equation}\label{eqn:lyapconj}
    \begin{aligned}
        L(E,\epsilon)&=L_A(E,\epsilon) -\frac1{2\pi}\int_0^{2\pi}\mathrm d\theta \,\log|t_0(\theta+\mathrm i\epsilon)|\\
        &=L_A(E,\epsilon)-\frac{p-1}{2}|\epsilon|,
    \end{aligned}
    \end{equation}
    where the integral follows from Jensen's formula. A naive attempt at applying the method of almost constant cocycles yields, as $\epsilon\rightarrow+\infty$,
    \begin{equation}
        \mathcal T_n(\theta+\mathrm i\epsilon)\longrightarrow
        \begin{pmatrix}0&-e^{\mathrm i\pi(p-1)\beta}\\1&0\end{pmatrix},
    \end{equation}
    which is unitary, so that $L(E,\epsilon)\rightarrow0$ and $L_A(E,\epsilon) = (p-1)|\epsilon|/2$ at asymptotically large $|\epsilon|$. Taking $\epsilon \rightarrow 0$ provides only $\gamma(E) \geq 0$.
    By the vanishing of the hopping coefficients along the real line, however, $L(E,\epsilon)$ fails to be convex at $\epsilon=0$, although it remains continuous there. The subtraction in Eq.~\eqref{eqn:lyapconj} produces a cusp, and $L(E,\epsilon)$ decreases from $\gamma(E)$ at $\epsilon=0$ to zero at large $|\epsilon|$, rather than increasing as it would for a nonsingular cocycle. The convexity of $L(E,\epsilon)$ away from the real line does hold, and it is possible to numerically resolve energy-dependent ``kinks'' where the magnitude of the slope of $L(E,\epsilon)$ decreases at finite $\epsilon$. It is the positions of these kinks, which are invisible at asymptotically large $\epsilon$, that set $\gamma(E)$.
    The off-diagonal structure of the transfer matrices in Eq.~\eqref{eqn:1deff} is thus such that complexification arguments do not provide a mechanism for the energy dependence of the Lyapunov exponent.
    A related failure of the conclusions of the global theory for a singular cocycle is noted in Ref.~\cite{jitomirskaya_arithmetic_2017} for the Maryland model. We regard the present models as examples demonstrating that the assumption that the hopping coefficients are bounded away from zero on a complex strip enclosing the real line is necessary.

   \begin{figure*}[tb]
        \centering
        \includegraphics[width=.7\linewidth]{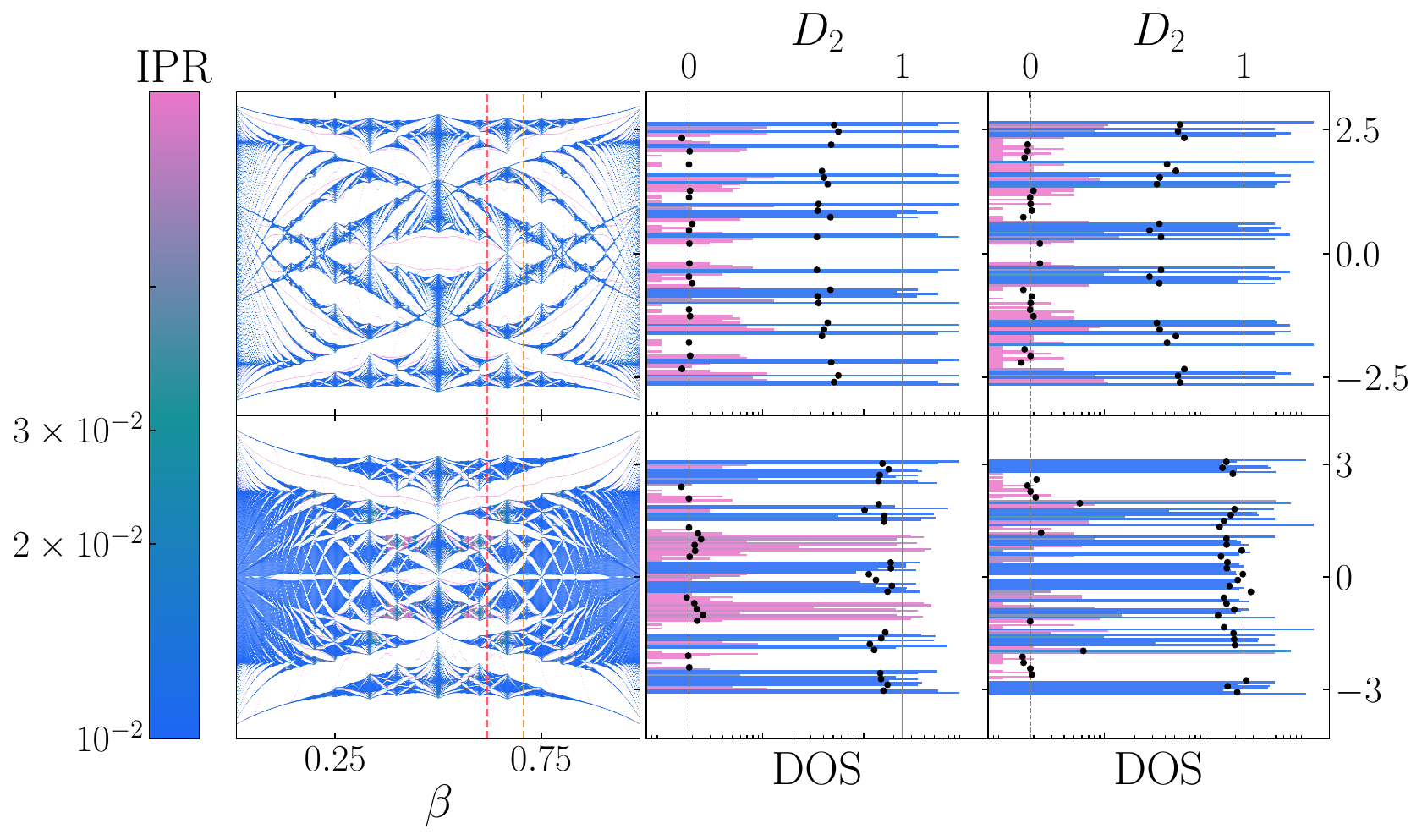}
        \caption{The Hofstadter butterfly for the square (top) and triangular (bottom) ladder lattices at $L = 500$ unit cells ($N = 1000$ sites). The colored vertical lines correspond to $\beta_1 = 1/\varphi$ and $\beta_2 = 1/\sqrt{2}$. 
        On the right are $\ipr$-colored densities of states below the energy-resolved fractal dimension $D_2$. 
        The data is computed at $L = 500$ and averaged over 20 values of $\theta$. 
        $D_2$ is computed as the slope of $\log\langle\ipr\rangle$ against $\log N$ for the same 20 values of $\theta$ for $L = 89, 144, \dots, 987$.
        }
        \label{fig:ladders}
    \end{figure*}

   \section{Positive Lyapunov exponent in the triangular pendant chain}\label{app:3stub}
   Here, we explicitly execute the Schur complement of the model specified in Fig.~\ref{fig:fig2} and, following the program of App.~\ref{app:complex}, produce a positive lower bound on the effective Schr\"odinger operator's Lyapunov exponent.
   To start, we write out the eigenvalue equation
   \begin{equation}
      E |\Psi_n\rangle = T_n ( |\Psi_{n+1}\rangle  +|\Psi_{n-1}\rangle) + V_n(\theta) |\Psi_n\rangle,
   \end{equation}
   with 
   \begin{equation}
       |\Psi_n \rangle = \begin{pmatrix}
           \psi_{3n} \\
           \psi_{3n+1} \\
           \psi_{3n+2} \\
       \end{pmatrix}.
   \end{equation}
   Essentially, we must solve 
   \begin{equation}
       E \begin{pmatrix}
           \psi_{3n+1} \\ 
           \psi_{3n+2}
       \end{pmatrix}
       =
      \begin{pmatrix}
          0 & 1 \\ 
          1 & 0 
      \end{pmatrix} 
        \begin{pmatrix}
           \psi_{3n+1} \\ 
           \psi_{3n+2}
       \end{pmatrix}
       + 
       \begin{pmatrix}
           e^{-\mathrm i \Phi_n/2} \\ 
           e^{\mathrm i \Phi_n/2}
       \end{pmatrix}
       \psi_{3n},
   \end{equation}
   yielding, for $E\neq\pm1$,
   \begin{equation}
   \begin{aligned}
       \begin{pmatrix}
           \psi_{3n+1} \\ 
           \psi_{3n+2} \\ 
       \end{pmatrix}
       = 
       \begin{pmatrix}
           E & -1 \\ 
           -1 & E
       \end{pmatrix}^{-1}
       \begin{pmatrix}
           e^{-\mathrm i \Phi_n/2} \\
           e^{\mathrm i \Phi_n/2} \\
       \end{pmatrix}\psi_{3n} \\ 
       = \frac{1}{E^2 - 1} \begin{pmatrix}
           E e^{-\mathrm i \Phi_n/2} + e^{\mathrm i \Phi_n/2} \\
           e^{-\mathrm i \Phi_n/2} + E e^{\mathrm i \Phi_n/2} \\
       \end{pmatrix}\psi_{3n}.
       \end{aligned}
   \end{equation}
   Hence, we arrive at
   \begin{equation}
       E\psi_{3n} = \psi_{3n + 3} + \psi_{3n - 3} +\frac{ 2E + 2 \cos(\Phi_n) }{E^2 - 1} \psi_{3n}.
   \end{equation}
   Organizing this into a transfer matrix, we find 
   \begin{equation}
       \mathcal T_n(\theta) = 
       \begin{pmatrix}
           E - \frac{ 2E + 2  \cos(2 \pi \beta n + \theta) }{E^2 - 1} & -1 \\ 
           1 & 0 
       \end{pmatrix}
   \end{equation}
   and 
   \begin{equation}
       \tilde{\mathcal{T}}_n(\theta,\epsilon) = e^{\epsilon}e^{-\mathrm i ( 2 \pi \beta n+ \theta)}\begin{pmatrix}
          -\frac{1}{E^2 - 1}  & 0 \\
          0 & 0 
       \end{pmatrix},
   \end{equation}
   so, at asymptotically large $\epsilon$, we have
   \begin{equation}
       L(E,\epsilon) = -\log |E^2 - 1| + |\epsilon|,
   \end{equation}
   which, by convexity, gives $\gamma(E)\geq\max(0,-\log|E^2-1|)$.
   For energies in the spectrum, where the cocycle cannot be uniformly hyperbolic, the quantization of the slope in $\epsilon$ extends this to all $\epsilon$ \cite{avila_global_2009}:
   \begin{equation}
       L(E,\epsilon) = \max(-\log |E^2 - 1| + |\epsilon|,0),
   \end{equation}
   which immediately leads to the desired result at $\epsilon = 0$, in agreement with the known Lyapunov exponent of the AAH model \cite{bourgain_jitomirskaya_2002}.
   Here, for $E\neq\pm1$, $E$ is in the spectrum of the full Hamiltonian if and only if $E-2E/(E^2-1)$ is in the spectrum of the AAH operator with unit hopping and potential $2\cos(\Phi_n)/(E^2-1)$.

Since the reduction to the AAH model holds energy by energy, its localization properties carry over, just as in other models with energy-dependent effective potentials \cite{Wang_2023}. For $|E|>\sqrt{2}$, the effective potential is subcritical, and the Hamiltonian has no normalizable eigenstates \cite{avila_ac_2008}. For $0<|E|<\sqrt{2}$ with $E\neq\pm1$, it is supercritical, and every eigenstate is exponentially localized with localization length $1/\gamma(E)$ for Diophantine $\beta$ and almost every $\theta$ \cite{jitomirskaya_metal_1999}. The energies $E=\pm\sqrt{2}$, at which the effective potential is critical, are thus exact mobility edges. At $E = 0$, the effective potential is again critical, and the localization length diverges as $1/E^{2}$ on approaching the center of the spectrum. 

   It is easy to generalize our model to longer pendant loops with variable pendant hopping $t$. 
   Indeed, for a 1D chain with $p\geq2$ pendant sites per loop, unit spine hopping, and hopping strength $t>0$ on all $p+1$ edges of each loop, we have, on the spectrum and away from the zeros of $U_p(E/(2t))$, 
   \begin{equation}
       \gamma(E) = \max\left(0, \log\left|\frac{t}{U_p\left(\frac{E}{2 t}\right)}\right|\right),
   \end{equation}
   where $U_p$ is the Chebyshev polynomial of the second kind. We remark that $U_{p-1}$ also appears in  Eq.~\eqref{eqn:1deff} as the nearest-neighbor hopping coefficients rather than the on-site potential. 

    Our arguments only apply when $\gamma(E)$ is finite. Complexification arguments of this kind do not apply cleanly at energies where $\gamma(E)$ diverges, which are precisely the eigenenergies of the pendant sites when detached from the main ``spine.'' Of course, this is no surprise: the denominator in the effective energy-dependent on-site potential is proportional to the characteristic polynomial of the Hamiltonian matrix restricted to the sites to be decimated.

    \section{Square and triangular ladder lattices}\label{sec:ladder}

    Unlike the other models we have studied in this paper, the square and triangular ladders cannot be neatly reduced to 1D chains by squaring or invoking the Schur complement. 
    The Hamiltonian blocks for these two lattices are given by 
    \begin{equation}
        \begin{aligned}
             T_n^{\text{sq.}}(\theta) &= 
             \begin{pmatrix}
                 e^{\mathrm i \Phi_n} & 0 \\ 
                 0 & 1
             \end{pmatrix}, \\ 
             V_n^{\text{sq.}}(\theta) &= 
             \begin{pmatrix}
                 0 & 1 \\ 
                 1 & 0 
             \end{pmatrix},
        \end{aligned}
    \end{equation}
    and 
    \begin{equation}
        \begin{aligned}
             T_n^{\text{tri.}}(\theta) &= 
             \begin{pmatrix}
                 e^{\mathrm i \Phi_{2n}} & 1 \\ 
                 0 & e^{-\mathrm i \Phi_{2n+1}}
             \end{pmatrix}, \\ 
             V_n^{\text{tri.}}(\theta) &= 
             \begin{pmatrix}
                 0 & 1 \\ 
                 1 & 0 
             \end{pmatrix}.
        \end{aligned}
    \end{equation}
    The triangular ladder has two plaquettes per unit cell and hence cell frequency $2\beta$. Both sets of intercell hopping matrices are nonsingular. From this fact, it follows that no application of magnetic fields, quasiperiodic or otherwise, can force the eigenstates to be strictly local. 
    Unlike for the models in Eq.~\eqref{eqn:pchain}, an absolutely continuous spectrum is not forbidden \textit{a priori} here. In the glued tree chains, it is excluded because the real zeros of the hopping coefficients in Eq.~\eqref{eqn:1deff} produce arbitrarily weak bonds infinitely often \cite{simon_trace_1989}. 

    As shown in Fig.~\ref{fig:ladders}, the square ladder lattice appears to host both critical and localized eigenstates with an energy- and $\beta$-dependent fractal dimension.
    Although the number of eigenstates with $D_2 \sim 0$ is small, these are not simply boundary modes.
    The triangular ladder, on the other hand, has mostly extended eigenstates, except for a small but appreciable set of localized eigenstates at some values of $\beta$. 

    Given even very uncomplicated lattices, it is unclear what should be regarded as typical behavior: it is plausible that criticality is less rare in the context of magnetic quasicrystals than in the context of quasiperiodic Schr\"odinger operators.

\section{Numerical results on few-body delocalization}\label{app:fewbody}

\begin{figure*}[t]
 \centering
 \includegraphics[width=\textwidth]{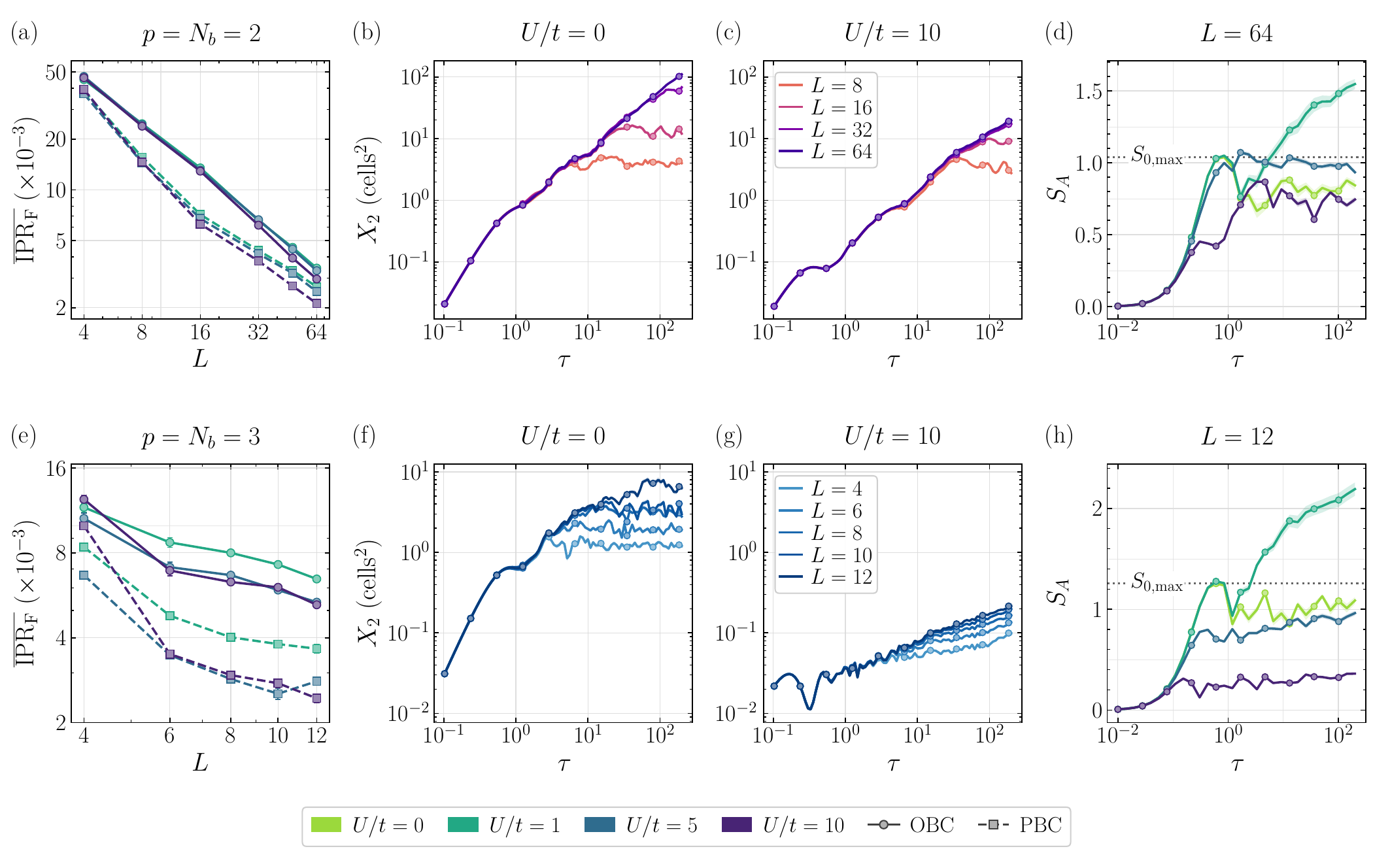}
 \caption{Occupation-basis participation, spatial expansion, and entanglement entropy for $p=N_b=2$ (top) and $3$ (bottom) glued tree chain lattices. (a,e) Mean Fock IPR for $U/t=1,5,10$, with open (solid circles) and periodic (dashed squares) boundaries. We order the energies as $E_0\leq\cdots\leq E_{\mathcal D-1}$ and then average over $\lfloor\mathcal D/4\rfloor\leq k<\lfloor3\mathcal D/4\rfloor$, retaining only levels whose two adjacent gaps exceed $\delta=10^{-10}\max\{t,|E_1|,|E_{\mathcal D-2}|\}$. This excludes unresolved degeneracies, whose individual eigenvector IPRs are basis dependent. State means are calculated separately for each phase and then averaged over $50$ phases. (b,c,f,g) Open-chain $X_2$ at $U/t=0,10$, averaged over at least $100$ phases. (d,h) Open-chain entanglement entropy at $L=64$ ($p=2$) and $L=12$ ($p=3$), averaged over $50$ phases; dotted lines correspond to Eq.~\eqref{eqn:condensateceiling}. }
 \label{fig:fewbodyfurther}
\end{figure*}

The calculations in this section use the Hamiltonian of Eq.~\eqref{eqn:interacting} with $N_b$\,$=$\,$p$\,$=$\,$2,3$ at $\beta$\,$=$\,$1/\varphi$, and every quantity is averaged over phases $\theta$ drawn uniformly from $[0,2\pi)$. We consider open chains that have $N$\,$=$\,$(p+1)L+1$ sites, as in Sec.~\ref{sec:interactions}, as well as periodic chains, with $N=(p+1)L$ sites. In the latter construction, the ring is closed by placing $e^{-\mathrm iW}T_{L-1}$ in block row $L-1$ and column $0$ of the Hamiltonian in Eq.~\eqref{eqn:blocktr}, and its Hermitian conjugate in the transposed position, where $W=(p-1)\sum_{n=0}^{L-1}\Phi_n/2$ modulo $2\pi$ sets the flux through the ring; all other blocks are those of Eq.~\eqref{eqn:pchain}.

A common way of determining whether a system exhibits a localized or delocalized spectrum is via the disorder-averaged statistics of its energy levels. Owing to the flat-band structure, however, these statistics are no longer meaningful here. From App.~\ref{app:flatband}, we know that, for $p\geq 3$, there are $p-2$ zero modes supported on the branches of each unit cell. If we consider $N_b$ such zero modes from $N_b$ distinct unit cells, their product does not have any double occupancy and is therefore annihilated by the interaction term. The hopping term also annihilates this state, which is thus an exact zero-energy eigenstate for every $U$. We find that there are $\binom{L}{N_b}(p-2)^{N_b}$ such states. A similar argument can be applied for $p=2$, where the zero modes span two neighboring unit cells: as long as there is no overlap between the supports of the zero modes, their product is again a zero-energy eigenstate. In reality, though, there are far more degeneracies. For example, taking $p=N_b=3$ and $L=12$ with open boundaries, we should have $220$ states from the above argument. Instead, $1481$ of the $\mathcal D=20825$ levels lie at zero energy. Furthermore, we find the same number of zero-energy levels for all values of $U$ considered (here, $U/t=1,5,10$). We also observe that, for the largest values of $L$, between $15\%$ and $23\%$ of the levels in the central half of the spectrum belong to numerically degenerate clusters. These clusters can be removed, and the adjacent-gap ratio of the remaining levels can be calculated. The result, however, depends on the cutoff chosen to define a degeneracy, the boundary conditions of the system, and the length of the chain. Consequently, no conclusions can be drawn from it.

The remaining eigenstates are well defined. We characterize them using their inverse participation ratio with respect to the occupation basis. To do so, we write a normalized eigenstate as $\lvert\Psi_k\rangle = \sum_{\boldsymbol n} c_{\boldsymbol n k} \lvert\boldsymbol n\rangle$, where $\boldsymbol n$ labels the $\mathcal D$ occupation-basis states with $\sum_x n_x=N_b$, and define
\begin{equation}\label{eqn:fockipr}
 \mathrm{IPR}_{\mathrm F}(k)=\sum_{\boldsymbol n}|c_{\boldsymbol n k}|^4,
 \qquad \mathcal D=\binom{N+N_b-1}{N_b}.
\end{equation}
The inverse of $\mathrm{IPR}_{\mathrm F}$ tells us how many configurations the state effectively occupies. The average $\mathrm{IPR}_{\mathrm F}$ for the nondegenerate levels in the central half of the spectrum (see the caption for exact details) is shown in Figs.~\ref{fig:fewbodyfurther}(a,e). For $p=2$, we find that this average decreases with increasing chain length for all values of the interaction strength. For example, at $U/t=1$, it drops from $0.0450$ to $0.00343$ as $L$ grows from $4$ to $64$ on the open chain. If the system had localized few-body states, then their $\mathrm{IPR}_{\mathrm F}$ would remain constant, as the number of occupied configurations would be fixed. However, we do not see this behavior. For $p=3$, the results are somewhat less clear. The average decreases only slowly with $L$, going, for instance, from $0.0116$ to $0.00646$ (open chain, $U/t=1$) between $L=4$ and $L=12$. Moreover, puzzlingly, it is not monotonic for periodic boundary conditions at $U/t=5$. We caution that, in both cases, $\mathcal D$ itself increases with $L$. Besides, the average is taken over the middle of the spectrum, meaning that the overlap between these eigenstates and the initial state in Sec.~\ref{sec:interactions} may be very small.

The few-body dynamics also lead to similar conclusions. We plot the second moment $X_2$ of Eq.~\eqref{eqn:dynobs} for various chain lengths in Figs.~\ref{fig:fewbodyfurther}(b,c,f,g). Although the curves for different $L$ overlap at early times, once the density reaches the ends of the shorter chain, its $X_2$ saturates, and the curves for the short and long chains separate. By $\tau=200$, the two noninteracting bosons have attained $X_2=68.1\pm1.6$ for $L=32$ and $110.6\pm3.1$ for $L=64$. The dynamical behavior is roughly the same for $U/t=10$, apart from the time-scale renormalization---the slower clock---described in Sec.~\ref{sec:interactions}. Hence, the plateaus seen in the short chains arise from finite-size effects rather than localization. Even for $p=3$, where the moment at $U/t=10$ is extremely small, $X_2$ is larger in longer chains, growing from $0.0987\pm0.0035$ for $L=4$ to $0.2166\pm0.0045$ for $L=12$. As noted in Sec.~\ref{sec:interactions}, however, $X_2$ also counts the bosons outside the on-site cluster, so this growth need not reflect motion of the triplon itself. We have not attempted to fit a transport exponent to these curves, as the slope of the fit is highly dependent on the fitting window.

Finally, it is natural to ask whether the interactions do anything more than slow down quasiparticle transport. To this end, it is useful to consider the entanglement entropy
\begin{equation}\label{eqn:fewbodyentropy}
 S^{}_A=-\operatorname{Tr}\rho^{}_A\log\rho^{}_A
\end{equation}
between the sites $A=\{0,\ldots,x_0\}$, from the start of the chain up to and including the initial hub, and the rest of the chain. Here, $\rho_A$ is the reduced density matrix of region $A$. If there are no interactions at all, then the entanglement is fundamentally limited, since all $N_b$ bosons occupy the same time-evolving orbital; that is, the state is $(\sum_xu^{\pdagger}_xb_x^\dagger)^{N_b}|0\rangle/\sqrt{N_b!}$, with $\sum_x|u^{\pdagger}_x|^2=1$. Hence, the probability of finding $k$ bosons in region $A$ follows a binomial distribution,
\begin{equation}\label{eqn:condensateschmidt}
 \lambda_k=\binom{N_b}{k}w^k(1-w)^{N_b-k},
 \quad w=\sum_{x\in A}|u_x|^2.
\end{equation}
These are the Schmidt probabilities of the state. For $N_b=2,3$, the entropy of this state becomes $N_bh(w)-N_bw(1-w)\log N_b$, where the usual binary entropy is $h(w)=-w\log w-(1-w)\log (1-w)$. The maximum of this function occurs at $w=1/2$, giving
\begin{equation}\label{eqn:condensateceiling}
 S_{0,\max}=N_b\log2-\frac{N_b}{4}\log N_b
 =\begin{cases}1.03972\ldots,&N_b=2,\\1.25548\ldots,&N_b=3.\end{cases}
\end{equation}
No matter how much the noninteracting bosons spread, their entanglement entropy can never exceed this bound. The interacting bosons, on the other hand, do reach a higher entropy. For $U/t=1$ and $\tau=200$, we see that the average entanglement  entropy is $1.547\pm0.035$ for $p=2$ and $2.192\pm0.066$ for $p=3$, as shown in Figs.~\ref{fig:fewbodyfurther}(d,h). Both of these values are still increasing with time. This tells us that, at weak coupling, certain correlations form that cannot be captured by one spreading orbital. At strong coupling ($U/t=10$), in contrast, the entropy stays below $S_{0,\max}$ for both systems (as expected, since the bosons should behave like a single heavy particle). Note also that, since there are only $N_b$ bosons in the system, there are at most $\binom{N_A+N_b}{N_b}$ occupation states in region $A$. Here, $N_A=x_0+1$, so the entropy is always less than or equal to $\log \binom{N_A+N_b}{N_b}=\mathcal{O}(\log N)$, regardless of whether the system is localized. Therefore, we gain no information from not seeing a volume-law growth.

\end{appendix}

\bibliography{refs}

\end{document}